\documentclass[journal,onecolumn]{IEEEtran}
\usepackage{epsfig}
\usepackage{amssymb}
\usepackage{amsthm}
\usepackage{cite}
\usepackage{citesort}
\usepackage[cmex10]{amsmath}
\usepackage{color}
\usepackage{epsfig}
\usepackage{url}
\usepackage{enumerate}
\usepackage{epstopdf}
\renewcommand{\baselinestretch}{0.96}

\newcommand*{\suchthat}{\;\ifnum\currentgrouptype=16 \middle\fi|\;}

\allowdisplaybreaks
\renewcommand{\baselinestretch}{0.96}
\newcounter{MYtempeqncnt}
\begin{document}
\title{State of Charge Evolution Equations for Flywheels}
\author{D. Fooladivanda, G. Mancini, S. Garg, and C. Rosenberg\\ 
Dept. of Electrical and Computer Engineering,
University of Waterloo, Canada\\ Email:\{dfooladi, gmancini, s6garg, cath\}@uwaterloo.ca \vspace{-0.05in}}

\maketitle

\section{Introduction}
Flywheel energy storage devices are composed of a spinning composite disk in an low-pressure enclosure
designed to contain the debris in the case of operation failure \cite{beaconpower}.
An electric motor-generator connector is used to convert the electric energy to kinetic energy by applying a torque on the flywheel.
The torque can be positive (charge) or negative (discharge). The flywheel levitates over magnetic bearings in order to reduce friction.
A cooling system must be used to remove the heat produced by the remaining friction and the electric components.
To reduce drag losses, a vacuum pump is used to create a low-pressure vacuum in the enclosure.
Flywheel technology has greatly improved in the past few years such that modern flywheels can rotate up to 100,000 round per minute (rpm) in
a low-pressure vacuum enclosure and achieve a very high energy density \cite{congressreport}.
The fast response time of flywheels makes this technology well suited for grid support (especially frequency regulation) \cite{deregulated}, \cite{probabilistic}, \cite{edison}.
Comprehensive overviews of flywheel energy storage devices are provided in \cite{fwreview}, \cite{fwreview2}.

Extensive work has been done on flywheel energy storage devices and their modeling, but most of these works rely on simulation and circuit models \cite{fwcomplex},\cite{fwdiesel}, \cite{fwacfaults},\cite{windpower}. Nassar \textit{et al.} \cite{activedistribution} propose a simple algorithm for simulating flywheel energy storage devices. However, this model does not include the charging/discharging inefficiencies and the self-discharge of the flywheel. Flywheels can suffer from significant leakage rates, due to frictional windage and magnetic losses from the bearings and motor-generator components \cite{losses1}, \cite{losses2}. Hearn \textit{et al.} \cite{optimalcontrol} make a conservative simplification on the core losses, and model the state of charge (SoC) evolution of a flywheel with a first-order differential equation.

A flywheel energy storage system based on a doubly-fed induction motor-generator is composed
of a wound-rotor induction machine and a cycloconverter. These storage devices are capable of both active and reactive power control while the conventional synchronous-speed rotary condenser is only capable of reactive power control. By adjusting the rotor speed of the doubly-fed induction machine, the motor-generator can either provide the electric power to the grid or draw it from the grid. Even though the model of Hearn \textit{et al.} \cite{optimalcontrol} includes a detailed description of the frictional windage and magnetic losses, it neglects the impact of the active power controller on the supplied/drawn energy from the flywheel. The purpose of this study is to obtain formulas for the SoC evolution of flywheels. We begin with the model proposed by Hearn \textit{et al.} \cite{optimalcontrol}, and additionally consider the impact of the active power controller on the SoC evolution of the flywheel energy storage.

\section{Flywheel Energy Storage}\label{flywheel_model_diff}
The energy stored in an energy storage device is mainly determined by the
charged/discharged energy and the storage losses. When the charge and discharge rates are sufficiently slow,
the charging and discharging efficiencies remain constant. In such cases, the time rate of change of stored energy $E$ in the energy storage device is proportional to the charging/discharging power multiplied by a factor. The simplest evolution model for the stored energy is as follows:
\begin{align}
&\label{SoC_storage}\frac{d E}{dt}=\eta_{\text{eff}}~P_{in}\\
&\label{SoC_storage_eff}\eta_{\text{eff}}=\left\{\begin{array}{cc}
   {e_c,} & {\text{if}~P_{in}\ge0} \\
   {e_d,} & {\text{if}~P_{in}<0} \\
\end{array}\right.
\end{align}
where $P_{in}$ and $\eta_{\text{eff}}$ denote the input power to the storage and the charging/discharging efficiency of the storage. The storage must supply power $\left(-P_{in}\right)$ if $P_{in}$ is negative, and must draw power $P_{in}$ from the grid if it is positive. The factor $\eta_{\text{eff}}$ can be represented by the charging and discharging efficiencies, $e_c$ and $e_d$, respectively. Clearly, the input power $P_{in}$ should be less than the rated power of the storage device.

\begin{figure}
\begin{center}
\includegraphics[width=3.2in]{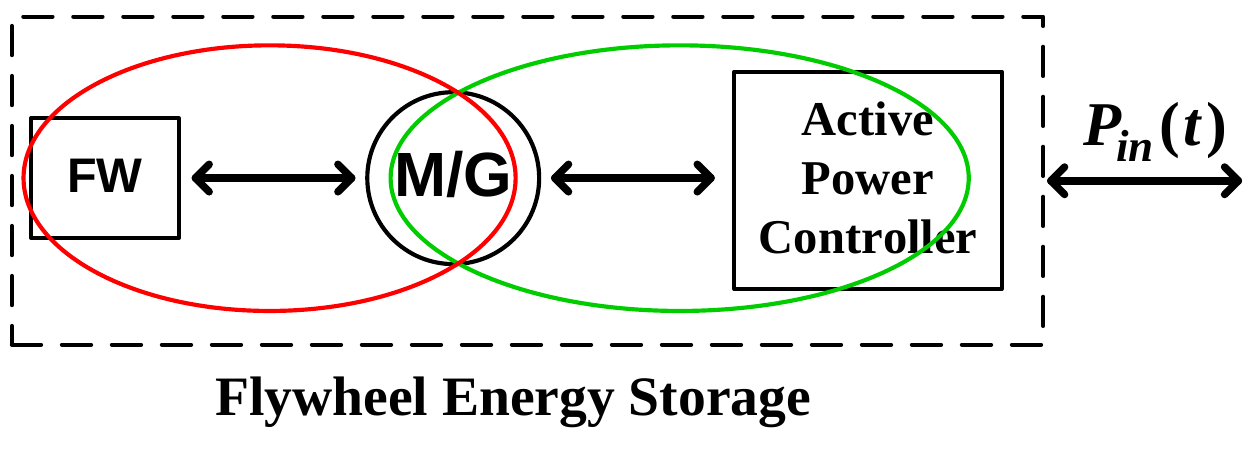}
\caption{A flywheel energy storage system comprising an induction machine, a flywheel, and an active power controller: $P_{in}(t)$ denotes the input power from the grid into the storage system. The system can be decomposed into two subsystems: the electrical system, and the mechanical system. These two subsystems are coupled together. The electrical subsystem is shown with a green ellipse while the mechanical system is shown with a red ellipse.}
\label{block_diag}
\end{center}
\end{figure}

The differential equation in (\ref{SoC_storage}) can describe the SoC evolution for electromechanical storage devices in which the charge and discharge processes are affected by negligible ohmic losses, and the storage can quickly convert electrical energy to mechanical energy, and vise versa. A flywheel energy storage system is composed of an induction machine, a flywheel, and an active power controller, as shown in Fig. \ref{block_diag}. Flywheels use motor-generators to electromechanically convert energy into and out of the flywheel. The conversion of electrical energy to mechanical energy, and vise versa, in motor-generators is not instantaneous and typically takes between 0.05 to 0.4 seconds
Moreover, frictional windage and magnetic losses from the bearing and motor-generator components are not negligible. Due to such non-idealities, the differential equation in (\ref{SoC_storage}) cannot accurately describe the stored energy evolution for flywheels.

We consider a flywheel energy storage system comprising an induction machine, a flywheel, and an active power controller, and decompose the system into the mechanical subsystem and the electrical subsystem, as shown in Fig. \ref{block_diag}. To model the SoC evolution for flywheels, we first focus on the mechanical subsystem composed of a flywheel and an induction machine, and model frictional windage and magnetic losses. We then focus on the electrical subsystem, and approximate the electrical subsystem with a first-order system with a certain time-constant.

\subsection{The Mechanical Subsystem}
Typically, motor-generators are highly efficient, i.e., they can operate at efficiencies higher than 90\% \cite{hebner}. Because of this, the energy losses of these storage devices are highly dependent on leakage rates which result from frictional windage and magnetic losses from the bearing and motor-generator components. The windage loss for a flywheel energy storage device is determined by the flywheel geometry, surface area, and chamber pressure. In \cite{losses1}, Liu \textit{et al.} show that the windage loss is proportional to the square speed of the flywheel $\omega$ multiplied by a constant factor if the flywheel is operating in a vacuum under slip or free molecular flow conditions. The constant factor is determined by the chamber pressure, flow regime, and geometry of the flywheel.

The standing losses from magnetic bearings and motor-generator components originate from lamination core losses. Using the Steinmetz equation, lamination core losses, $P_c$, which consist of magnetic hysteresis loss and eddy current loss, can be estimated by \cite{losses2}
\[
P_c=k_h f_e Q^n+k_e f_e^2 Q^2
\]
where $f_e$ is the electrical frequency of the external magnetic field, $Q$ is the peak magnetic flux density, $k_h$, $k_e$, and $n$ are
the coefficients which depend on the lamination material, thickness, and conductivity \cite{losses2}. Since the electrical frequency is proportional to the mechanical speed and the number of pole pairs in the magnetic bearing or motor-generator design, eddy current losses are proportional to the square of speed while hysteresis losses are linear with speed.

In \cite{optimalcontrol}, using a conservative simplification of the core losses, Hearn \textit{et al.} propose a simple evolution equation which takes into account windage and lamination losses.
The authors consider only the quadratic dependence on speed,
and define a loss factor $Q_{\text{loss}}$ which represents a
summation of frictional and magnetic losses due to windage and lamination losses.
By doing this, the time rate change of stored energy in the flywheel due to windage and lamination losses can be represented by
\[
\frac{d E}{dt}=-Q_{\text{loss}} \omega^2~.
\]
Given a speed $\omega$, the stored energy $E$ in a flywheel with inertia $J$ is equal to $\frac{1}{2} J \omega^2$. Therefore, the time rate change of stored energy in the flywheel due to windage and lamination losses can be represented by a first-order differential equation as follows:
\[
\frac{d E}{dt}=-\frac{1}{T_{\text{loss}}} E
\]
where $T_{\text{loss}}=\frac{J}{2 Q_{\text{loss}}}$ the time constant of the storage losses.

Let $E_{int}$ denote the initial energy of the flywheel at $t=0$ (i.e., $E(0)=E_{int}$). Using the first-order approximation of flywheel losses, the evolution of the stored energy in the flywheel can be represented by
\begin{align}
\label{SoC_flywheel}&\frac{d E(t)}{dt}=P_{in}^m(t)-\frac{1}{T_{\text{loss}}} E(t)\\
\label{initial}&E(0)=E_{int}
\end{align}
where $E(t)$ and $P_{in}^m(t)$ denote the kinetic energy stored in the flywheel energy storage and the mechanical input power to the flywheel at time $t$, respectively. Note that this differential equation can describe the  evolution of the stored energy in the flywheel when the stored energy $E(t)$ is less than or equal to the storage capacity.

Using Laplace transforms and the SoC evolution model of the mechanical subsystem, we can show that the stored energy in the mechanical subsystem at time instant $t$ can be computed by
\begin{equation}
\label{soc_exact}E(t)=E_{int} ~e^{\frac{-t}{T_{loss}}}+\int_{0}^{t}{{P_{in}^m(\tau)} ~e^{\frac{-(t-\tau)}{T_{loss}}}  d\tau}~.
\end{equation}
The evolution equation above enables us to compute the stored energy in the mechanical subsystem at each time instant $t$ for any power profile $P_{in}^m(t)$. The proposed model enables designers to study the effects of windage and lamination losses on the flywheel performance in practical systems.

\begin{figure*}[!t]
\normalsize
\setcounter{MYtempeqncnt}{\value{equation}}
\setcounter{equation}{8}
\begin{align}\small
\label{soc_fw}\overline{E_{k}}&=\left\{\begin{array}{cc}
   {\overline{G}{(P_{in},P_{prev})},} & {\textbf{if}~{P_{in} P_{prev}< 0}~~\textbf{and}~~{\frac{P_{in}}{P_{in}-P_{prev}}> e^{\frac{-\delta}{T_{cont}}}}} \\
   {\overline{H}{(P_{in},P_{prev})},} & {\textbf{otherwise}} \\
\end{array}\right.\\
\overline{G}{(P_{in},P_{prev})}&=\Gamma\left[P (\eta_{prev}-\eta_{in})  P_{in} {\left(\frac{P_{in}-P_{prev}}{P_{in}}\right)}^{\frac{T_{cont}}{T_{loss}}}+P_{in} T_{loss} \left(\Gamma^{-1}\eta_{in}-\eta_{prev}\right)-\left(\frac{P_{in}-P_{prev}}{Q}\right) \left(e^{\delta Q} \eta_{in}-\eta_{prev}\right)\right]\nonumber\\
\overline{H}{(P_{in},P_{prev})}&=\Gamma \eta_{\text{eff}} \left[P_{in} T_{loss} \left(\Gamma^{-1}-1\right)-\left(\frac{P_{in}-P_{prev}}{Q}\right) \left(e^{\delta Q}-1\right)\right]\nonumber\\
\setcounter{equation}{9}
\label{soc_app1}\widehat{E_{k}}&=\left\{\begin{array}{cc}
   {\widehat{G}{(P_{in},P_{prev})},} & {\textbf{if}~{P_{in} P_{prev}< 0}~~\textbf{and}~~{\frac{P_{in}}{P_{in}-P_{prev}}> e^{\frac{-\delta}{T_{cont}}}}} \\
   {\widehat{H}(P_{in},P_{prev}),} & {\textbf{otherwise}} \\
\end{array}\right.
\end{align}
\setcounter{equation}{\value{MYtempeqncnt}}
\hrulefill
\vspace*{4pt}
\end{figure*}

\subsection{The Electrical Subsystem}
A flywheel energy storage system based on a doubly-fed induction motor-generator is composed of a wound-rotor induction machine and a cycloconverter. By adjusting the rotor speed, the flywheel can either release the kinetic energy to the grid or absorb energy from the grid. Therefore, the motor-generator component has the capability of achieving not only reactive power control, but also active power control. In the literature, several power control strategies have been proposed for doubly-fed wound rotor induction machines \cite{ref6},\cite{ref7},\cite{ref8},\cite{ref9},\cite{akagi}. In \cite{akagi}, Akagi \textit{et. al} propose a control strategy which enables the induction motor-generator to perform active power control independent of reactive power control. The authors show that the transfer function of such controllers can be represented by a first-order system with a certain time-constant \cite{akagi}, representing the inertia of the conversion system which is
typically
between 0.05 to 0.4 seconds.
The speed of electrical power response in the motor-generator depends on the inductance of the electrical machine and the control gains.
We use the same approach, and model the electrical subsystem with a first-order system whose time-constant depends on the inductance of the electrical machine and the control gains of the active power controller.

Let $T_{cont}$ denote the time-constant of the combination of the active power controller and the induction machine. The output of the electrical subsystem $P_{in}^m(t)$, which is the input power of the mechanical subsystem (i.e., the flywheel and motor-generator), can be computed by\footnote{The symbols $``\star"$ and ``$\times$" represent the convolution and the dot product, respectively.}
\begin{align}
\label{first_order}&P_{in}^m(t)=\eta_{\text{eff}}(t)\times\left(P_{in}(t) \star H_{cont}(t)\right)\\
\label{power_conv}&H_{cont}(t)=\frac{1}{T_{cont}} ~e^{\frac{-t}{T_{cont}}}~,
\end{align}
where $P_{in}(t)$ and $H_{cont}(t)$ denote the input power into the energy storage system and the transfer function of the electrical subsystem (i.e., the combination of the active power controller and the induction machine), respectively. The charging/discharging efficiency $\eta_{\text{eff}}(t)$ represents the charging efficiency $e_c$ and the discharging efficiency $e_d$.

The proposed models for the mechanical and electrical subsystems enable us to compute the stored energy in the storage system at each time instant $t$ for any power profile $P_{in}(t)$. Next, we consider a constant power scenario, and derive the evolution equations for the stored energy in the storage system.

\section{Energy Evolution Equations}\label{flywheel_model_soc}
Let us assume that the time is slotted in time slot of size $\delta$, and that a constant power is applied for the duration of a time slot, i.e., the input power $P_{in}(t)$ is constant during each interval $[k\delta,(k+1)\delta]$ where $k\in\{0,1,2,\cdots\}$. Let $\Gamma$ denote the self-discharge efficiency of the storage. $\Gamma$ accounts for the losses after $\delta$ units of time, and it is equal to $e^{\frac{-\delta}{T_{loss}}}$. Under these assumptions, the stored energy at $t_k=k\delta$ can be computed by
\begin{align}\small
\label{soc_const_power}E(t_k)&=~\Gamma \times E\left(t_{k-1}\right)+\overline{E_{k}}
\end{align}
where $\overline{E_{k}}$ denotes the absorbed/supplied energy in time slot $k$. To compute $\overline{E_{k}}$, let $P_{in}$ and $P_{prev}$ denote the input power in time slot $k$ and time slot $k-1$, respectively. The value of $\overline{E_{k}}$ is given by (\ref{soc_fw}) at the top of the page. A sketch of the proof is provided in the appendix. The constants $\eta_{in}$, $\eta_{prev}$, and $\eta_{\text{eff}}$ are computed by
\begin{align}
\eta_{in}&=(e_c \mathbf{1}_{P_{in}\ge0} + e_d \mathbf{1}_{P_{in}<0})\nonumber\\
\eta_{prev}&=(e_c \mathbf{1}_{P_{prev}\ge0} + e_d \mathbf{1}_{P_{prev}<0})\nonumber\\
\eta_{\text{eff}} &=\left(\eta_{in} \mathbf{1}_{|P_{in}|> |P_{prev}|} + \eta_{prev} \mathbf{1}_{|P_{in}|\le |P_{prev}|}\right)\nonumber
\end{align}
$P$ and $Q$ are the coefficients which depend on the system parameters $T_{cont}$ and $T_{loss}$ as follows:
\begin{align}
P &= T_{loss}-\frac{T_{loss} T_{cont}}{T_{cont}-T_{loss}},~~Q=\frac{T_{cont}-T_{loss}}{T_{loss} T_{cont}}~.\nonumber
\end{align}

\addtocounter{equation}{2}

Our analytical results show that the stored energy in the storage system at time $t_k=k\delta$ is a function of the stored energy at $t_{k-1}$, the input power at $t_k$ (i.e., $P_{in}$), and the input power at $t_{k-1}$ (i.e., $P_{prev}$). In contrast to the conventional SoC evolution equations, the SoC of the flywheel storage system is not only a function of the SoC at $t_{k-1}$ and the input power at $t_k$, but also the input power at $t_{k-1}$. The results show that the dependence on $P_{in}$ and $P_{prev}$ is complex so that it cannot be neglected. Next, we focus on a constant power profile, and propose a method to approximately compute the stored energy in the flywheel energy storage system.

\section{Approximate Method}
The stored energy at a give time instant $t_k=k\delta$ is a function of the stored energy at $t_{k-1}$ and the supplied/drawn energy in time slot $k$. Typically, the system parameter $T_{loss}$ is very large compared to the time slot duration $\delta$. In such scenarios, the impact of the system losses on the supplied/drawn energy in time slot $k$ is negligible. We neglect the impact of the system losses on the absorbed/supplied energy in time slot $k$, and approximate the SoC at $t_k=k\delta$ as follows:
\begin{align}
E_{fw}^{App}(t_k)&=e^{\frac{-\delta}{T_{loss}}} E_{fw}^{App}(t_{k-1})+\widehat{E_k}\nonumber\\
\label{approx1_int1} \widehat{E_{k}}&=\int_{t_{k-1}}^{t_k}{{P_{in}^m(\tau)}~d\tau}
\end{align}
where $E_{fw}^{App}(t_{k})$ denotes the approximated energy at $t=t_{k}$. Clearly, we have $E_{fw}^{App}(0)=E_{int}$.

To compute $\widehat{E_{k}}$, let $P_{in}$ and $P_{prev}$ denote the input power in time slot $k$ and time slot $k-1$, respectively. The value of $\widehat{E_{k}}$ can be computed by (\ref{soc_app1}) at the top of the previous page.
\begin{align}\small
\widehat{G}(P_{in},P_{prev})&=\eta_{prev} \left[P_{in} T_{change}+P_{prev} T_{cont} \right]+\eta_{in} \left[P_{in} \left(\delta-T_{change}-T_{cont}\right)\right]\nonumber\\
\widehat{H}(P_{in},P_{prev})&=\eta_{\text{eff}} \left[P_{in}(\delta- T_{cont})+P_{prev} T_{cont}\right]~.\nonumber
\end{align}
where $T_{change}=-T_{cont}\text{Ln}\left(\frac{P_{in}}{P_{in}-P_{prev}}\right)$. Clearly, there is a gap between the real value of the stored energy and the approximated energy at each time slot. The following result enables us to compute the difference between the real value of the stored energy and the approximated value at each time slot. A sketch of the proof is provided in the appendix.

\addtocounter{equation}{1}

\textbf{Result 1:} Given $k$, $\delta$, $T_{loss}$, $T_{cont}$, and the power profile $P_{in}(t)$, we have:
\begin{align}
\label{gap_app2} \left|E(t_{k})-E_{fw}^{App}(t_{k})\right| \le \left(1-e^{\frac{-(k+1)\delta}{T_{loss}}}\right) R_{max} \delta
\end{align}
where $R_{max}=e_d \max_{0 \le t\le t_{k}} {\left|P_{in}(t)\right|}$. The main messages of the bound in (\ref{gap_app2}) can be summarized as follows:
\begin{itemize}
\item The proposed upper bound is increasing in $k$ since as $k \rightarrow \infty$, we have $e^{\frac{-(k+1)\delta}{T_{loss}}} \rightarrow 0$ . Therefore, we have
 \[\left|E(t_k)-E_{fw}^{App}(t_k)\right| \le R_{max}~\delta,~ \forall k~.\]
\item The approximate method will perform better if $\delta << T_{loss}$. Regardless of the value of $\delta$, the proposed model will follow the stored energy in the system since the gap between the real value of the stored energy and the approximated value is less than $R_{max} \delta$.
\item The worst power profile for this approximation is $P_{in}(t)=P_{rated}$ for all $t$.
\end{itemize}

\appendices
\section{Energy Evolution Equation}\label{app1}
Let us assume that the time is slotted in time slot of size $\delta$. At time instant $t_k=k\delta$, the stored energy in the system can be computed by
\begin{align}\small
E(t_k)&=E_{int} e^{\frac{-t_k}{T_{loss}}}+\int_{0}^{t_k}{{P_{in}^m(\tau)} e^{\frac{-(t_k-\tau)}{T_{loss}}}  d\tau}\nonumber\\
&=\left(E_{int} e^{\frac{-t_{k-1}}{T_{loss}}}\right) e^{\frac{-\delta}{T_{loss}}}+\int_{0}^{t_{k-1}}{{P_{in}^m(\tau)} e^{\frac{-(t_{k-1}+\delta-\tau)}{T_{loss}}}  d\tau}\nonumber\\
&+\int_{t_{k-1}}^{t_k}{{P_{in}^m(\tau)} e^{\frac{-(t_{k-1}+\delta-\tau)}{T_{loss}}}  d\tau}\nonumber\\
&=e^{\frac{-\delta}{T_{loss}}} \left(E_{int} e^{\frac{-t_{k-1}}{T_{loss}}} +\int_{0}^{t_{k-1}}{{P_{in}^m(\tau)} e^{\frac{-(t_{k-1}-\tau)}{T_{loss}}}  d\tau}\right)\nonumber\\
&+\int_{t_{k-1}}^{t_k}{{P_{in}^m(\tau)} e^{\frac{-(t_k-\tau)}{T_{loss}}}  d\tau}\nonumber\\
\label{soc_main}&=e^{\frac{-\delta}{T_{loss}}} \times E(t_{k-1})+\int_{t_{k-1}}^{t_k}{{P_{in}^m(\tau)} e^{\frac{-(t_k-\tau)}{T_{loss}}}  d\tau}
\end{align}
where $P_{in}^m(t)=\eta_{\text{eff}}(t)\times\left(P_{in}(t) \star H_{cont}(t)\right)$ and $H_{cont}(t)=\frac{1}{T_{cont}} e^{\frac{-t}{T_{cont}}}$.

To compute $E(t_{k})$, define the absored/supplied energy $\overline{E}_k$ as follows:
\begin{align}
\label{energy_main}&\overline{E_{k}}=\int_{t_{k-1}}^{t_k}{{P_{in}^m(\tau)} e^{\frac{-(t_k-\tau)}{T_{loss}}} d\tau}~
\end{align}
where $P_{in}^m(t)$ is the input power of the mechanical subsystem. We assume that a constant power is applied for the duration of a time slot, i.e., the input power $P_{in}(t)$ is constant during each interval $[k\delta,(k+1)\delta]$ where $k\in\{0,1,2,\cdots\}$. We can show that:
\begin{align}\small
&\label{E_integral1}\overline{E_{k}}=\int_{0}^{\delta}{{\widehat{P_{in}^m(\tau)}} ~e^{\frac{-(\delta-\tau)}{T_{loss}}} d\tau}\\
\label{P_s1} &\widehat{P_{in}^m(t)}=\widehat{\eta_{\text{eff}}(t)}\left[P_{prev}+(P_{in}-P_{prev})(1-~e^{\frac{-t}{T_{cont}}})\right],
\end{align}
where $P_{in}$ and $P_{prev}$ denote the input power in time slot $k$ and time slot $(k-1)$, respectively. $\widehat{\eta_{\text{eff}}(t)}$ is qual to $e_c$ if $\widehat{P_{in}^m(t)}\ge0$; otherwise, it is equal to $e_d$.

Note that $\widehat{\eta_{\text{eff}}(t)}$ is a function of $\widehat{P_{in}^m(t)}$. More precisely, for each value of $t\in[0,\delta]$, the value of $\widehat{\eta_{\text{eff}}(t)}$ depends on the sign of $\widehat{P_{in}^m(t)}$. To simplify the term under the integral in (\ref{E_integral1}), we consider the following cases:
\begin{enumerate}
\item $\mathbf{P_{in}=0}$: If $P_{prev}\ge0$, $\widehat{P_{in}^m(t)}$ will be non-negative; otherwise, it is negative. Therefore, the value of $\widehat{\eta_{\text{eff}}(t)}$ is equal to $\eta_{prev}=(e_c \mathbf{1}_{P_{prev}\ge0} + e_d \mathbf{1}_{P_{prev}<0})$.
\item $\mathbf{P_{prev}= 0}$: If $P_{in}\ge0$, $\widehat{P_{in}^m(t)}$ will be non-negative; otherwise, it is negative. Therefore, the value of $\widehat{\eta_{\text{eff}}(t)}$ is equal to $\eta_{in}=(e_c \mathbf{1}_{P_{in}\ge0} + e_d \mathbf{1}_{P_{in}<0})$.
\item $\mathbf{P_{in} P_{prev}> 0}$: In this case, $P_{in}$ and $P_{prev}$ have the same sign. It can be verified that $\widehat{P_{in}^m(t)}$ has the same sign as $P_{in}$ and $P_{prev}$. Therefore, the value of $\widehat{\eta_{\text{eff}}(t)}$ is equal to $\eta_{in}=\eta_{prev}$.

\item $\mathbf{P_{in} P_{prev}< 0}$ and $\mathbf{\frac{P_{in}}{P_{in}-P_{prev}}\le e^{\frac{-\delta}{T_{cont}}}}$: Let us assume that $P_{in}>0$ and $P_{prev}<0$. In this case, the flywheel will be discharging during the first time instants before it begins charging. Hence, there exists a time instant (we call it $T_{change}$) so that $\widehat{\eta_{\text{eff}}(t)}=e_d$ for $0\le t \le T_{change}$, and $\widehat{\eta_{\text{eff}}(t)}=e_c$ for $T_{change} \le t \le \delta$. Using (\ref{P_s1}), we can show that $T_{change}=-T_{cont}\text{Ln}\left(\frac{P_{in}}{P_{in}-P_{prev}}\right)$. Since $\mathbf{\frac{P_{in}}{P_{in}-P_{prev}}\le e^{\frac{-\delta}{T_{cont}}}}$, we have $T_{change}\ge \delta$, i.e., $\widehat{P_{in}^m(t)}$ and $P_{prev}$ have the same sign over the time slot. Therefore, the value of $\widehat{\eta_{\text{eff}}(t)}$ is equal to $\eta_{prev}=(e_c \mathbf{1}_{P_{prev}\ge0} + e_d \mathbf{1}_{P_{prev}<0})$. Since $\mathbf{\frac{P_{in}}{P_{in}-P_{prev}}\le e^{\frac{-\delta}{T_{cont}}}}$, we have ${|P_{in}|\le |P_{prev}|}$. Therefore, the value of $\widehat{\eta_{\text{eff}}(t)}$ is equal to $\eta_{\text{eff}}=\left(\eta_{in} \mathbf{1}_{|P_{in}|> |P_{prev}|} + \eta_{prev} \mathbf{1}_{|P_{in}|\le |P_{prev}|}\right)$.
    
\item $\mathbf{P_{in} P_{prev}< 0}$ and $\mathbf{\frac{P_{in}}{P_{in}-P_{prev}}> e^{\frac{-\delta}{T_{cont}}}}$: In this case, we have $T_{change}< \delta$ since $\mathbf{\frac{P_{in}}{P_{in}-P_{prev}}>e^{\frac{-\delta}{T_{cont}}}}$. If $P_{in}>0$ and $P_{prev}<0$, there exists a time instant (we call it $T_{change}$) so that $\widehat{\eta_{\text{eff}}(t)}=e_d$ for $0\le t \le T_{change}$, and $\widehat{\eta_{\text{eff}}(t)}=e_c$ for $T_{change} \le t \le \delta$. Similarly, if $P_{in}<0$ and $P_{prev}>0$, we will have $\widehat{\eta_{\text{eff}}(t)}=e_c$ for $0\le t \le T_{change}$, and $\widehat{\eta_{\text{eff}}(t)}=e_d$ for $T_{change} \le t \le \delta$.
\end{enumerate}
The results above show that for the first four cases, $\widehat{\eta_{\text{eff}}(t)}$ is constant while for the last case, $\widehat{\eta_{\text{eff}}(t)}$ is a function of $t$.

For the first four cases, $\overline{E_{k}}$ can be computed by
\begin{align}\small
\overline{E_{k}}&=\Gamma \eta_{\text{eff}} \left[P_{in} T_{loss} \left(\Gamma^{-1}-1\right)-\left(\frac{P_{in}-P_{prev}}{Q}\right) \left(e^{\delta Q}-1\right)\right]\nonumber
\end{align}
where $\Gamma=e^{\frac{-\delta}{T_{loss}}}$. For the last case, $\overline{E_{k}}$ can be computed by
\begin{align}
\overline{E_{k}}&=\Gamma\left[P (\eta_{prev}-\eta_{in})  P_{in} {\left(\frac{P_{in}-P_{prev}}{P_{in}}\right)}^{\frac{T_{cont}}{T_{loss}}}+P_{in} T_{loss} \left(\Gamma^{-1}\eta_{in}-\eta_{prev}\right)-\left(\frac{P_{in}-P_{prev}}{Q}\right) \left(e^{\delta Q} \eta_{in}-\eta_{prev}\right)\right]~.\nonumber
\end{align}
$P$ and $Q$ are the coefficients which depend on the system parameters $T_{cont}$ and $T_{loss}$ as follows:
\begin{align}
P &= T_{loss}-\frac{T_{loss} T_{cont}}{T_{cont}-T_{loss}},~~Q=\frac{T_{cont}-T_{loss}}{T_{loss} T_{cont}}~.\nonumber
\end{align}
This completes the derivation of the energy evolution equation for the flywheel energy storage system.\qed

\section{Approximation}
We compute the SoC at $t_k=k\delta$ by
\begin{align}
E_{fw}^{App}(t_k)&=e^{\frac{-\delta}{T_{loss}}} E_{fw}^{App}(t_{k-1})+\widehat{E_k}\nonumber\\
\label{energy_app}\widehat{E_{k}}&=\int_{t_{k-1}}^{t_k}{{P_{in}^m(\tau)}~d\tau}
\end{align}
where $P_{in}^m(t)=\eta_\text{eff}(\tau)\times \left(P_{in}(t)\star H_{cont}(t)\right)$ is the input power of the mechanical subsystem. We assume that a constant power is applied for the duration of a time slot, i.e., the input power $P_{in}(t)$ is constant during each interval $[k\delta,(k+1)\delta]$ where $k\in\{0,1,2,\cdots\}$. We can show that $\widehat{E_{k}}=\int_{0}^{\delta}{{\widehat{P_{in}^m(\tau)}} ~ d\tau}$ where $\widehat{P_{in}^m(t)}$ is given by (\ref{P_s1}). Note that we can compute the value of $\widehat{\eta_{\text{eff}}(t)}$ by using the results in Appendix \ref{app1}.

For the first four cases, $\widehat{E_{k}}$ can be computed by
\begin{align}\small
\widehat{E_{k}}&= \eta_{\text{eff}} \left[P_{in}\delta+P_{in} T_{cont} \left(e^{\frac{-\delta}{T_{cont}}}-1\right)+P_{prev} T_{cont} \left(1-e^{\frac{-\delta}{T_{cont}}}\right)\right]\nonumber
\end{align}
For the last case, $\widehat{E_{k}}$ can be computed by
\begin{align}
\widehat{E_{k}}&=\eta_{prev} \left[P_{in} T_{change}+P_{prev} T_{cont} \right]+\eta_{in} \left[P_{in} \left(\delta-T_{change}-T_{cont}\right)+(P_{in}-P_{prev}) T_{cont} e^{\frac{-\delta}{T_{cont}}} \right]~.\nonumber
\end{align}

Typically, the time constant $T_{cont}$ is very small (i.e., $\delta >> T_{cont}$), and hence $e^{\frac{-\delta}{T_{cont}}}\rightarrow 0$.
For the first four cases, $\widehat{E_{k}}$ can be computed by
\begin{align}\small
\widehat{E_{k}}&= \eta_{\text{eff}} \left[P_{in}(\delta- T_{cont})+P_{prev} T_{cont}\right]\nonumber~.
\end{align}
For the last case, $\widehat{E_{k}}$ can be computed by
\begin{align}
\widehat{E_{k}}&=\eta_{prev} \left[P_{in} T_{change}+P_{prev} T_{cont} \right]+\eta_{in} \left[P_{in} \left(\delta-T_{change}-T_{cont}\right)\right]~.\nonumber
\end{align}
This completes the derivation of the energy evolution equation.\qed

\section{Results I}
Let us assume that the time is slotted in time slot of size $\delta$. Using (\ref{soc_main}), the stored energy in the system at time instant $t_k=k\delta$ can be written as follows:
\begin{align}
E(t_k)&=\sum_{i=0}^k \overline{E_{i}} e^{\frac{-(k-i)\delta}{T_{loss}}}\nonumber
\end{align}
where $\overline{E_{0}}=E_{int}$, and $\overline{E_{i}}$ can be computed by (\ref{energy_main}). Similarly, we can show that:
\begin{align}
E_{fw}^{App}(t_k)&=\sum_{i=0}^k \widehat{E_{i}} e^{\frac{-(k-i)\delta}{T_{loss}}}\nonumber
\end{align}
where $\widehat{E_{0}}=E_{int}$, and $\widehat{E_{i}}$ can be computed by (\ref{energy_app}).

Our goal is to compute the difference between the real value of the stored energy and the approximated value at $t=t_k$. We have:
\begin{align}
\left|E(t_k)-E_{fw}^{App}(t_k)\right|&=\left|\sum_{i=0}^k \left(\overline{E_{i}}-\widehat{E_{i}}\right) e^{\frac{-(k-i)\delta}{T_{loss}}}\right|\nonumber\\
\label{eq_app_1}&\le \sum_{i=0}^k \left|\left(\overline{E_{i}}-\widehat{E_{i}}\right)\right| e^{\frac{-(k-i)\delta}{T_{loss}}}~.
\end{align}
Therefore, to compute $|E(t_k)-E_{fw}^{App}(t_k)|$, we need to compute $|\overline{E_{i}}-\widehat{E_{i}}|$ for all $i\in\{0,1,\cdots,k\}$. Using the definitions of $\overline{E_{i}}$ and $\widehat{E_{i}}$, we can write:
\begin{align}
 \left|\overline{E_{i}}-\widehat{E_{i}}\right| &= \left|\int_{t_{i-1}}^{t_i}{{P_{in}^m(\tau) } ~e^{\frac{-(t_i-\tau)}{T_{loss}}} d\tau}-\int_{t_{i-1}}^{t_i}{{P_{in}^m(\tau)} ~ d\tau}\right|\nonumber\\
&\le \int_{t_{i-1}}^{t_i}{\left|{ P_{in}^m(\tau) } ~\left(e^{\frac{-(t_i-\tau)}{T_{loss}}}-1\right)\right| d\tau}\nonumber\\
&=\int_{t_{i-1}}^{t_i}{\left|{ P_{in}^m(\tau) }\right| ~\left(1-e^{\frac{-(t_i-\tau)}{T_{loss}}}\right) d\tau}\nonumber~.
\end{align}

Using the convolution properties, we can show that $\left|P_{in}^m(t)\right| \le \eta_{\text{eff}}(t) \left|P_{in}(t)\right|$ for all $t\in[0,t_{k}]$. Therefore, we have $\left|P_{in}^m(t)\right| \le \max{\{e_c,e_d\}} \left|P_{in}(t)\right|$. Typically, we have $e_c\le e_d$. Let $R_{max}=e_d \max_{0 \le t\le t_{k}} {\left|P_{in}(t)\right|}$. Since $\max_{t\in[t_{i-1},t_i]}{\left(1-e^{\frac{-(t_i-\tau)}{T_{loss}}}\right)}=(1-e^{\frac{-\delta}{T_{loss}}})$ , we have:
\begin{align}
 \left|\overline{E_{i}}-\widehat{E_{i}}\right| &\le \int_{t_{i-1}}^{t_i}{R_{max} ~(1-e^{\frac{-\delta}{T_{loss}}}) d\tau}\nonumber\\
\label{eq_app_2} &= R_{max} \delta(1-e^{\frac{-\delta}{T_{loss}}})~.
\end{align}
Now, by using (\ref{eq_app_1}) and (\ref{eq_app_2}), we have
\begin{align}
\left|E(t_k)-E_{fw}^{App}(t_k)\right|& \le R_{max}~(1-\Gamma^{k+1})\nonumber
\end{align}
where $\Gamma=e^{\frac{-\delta}{T_{loss}}}$. Note that $\Gamma^{k+1} \rightarrow 0$ as $k \rightarrow \infty$. Therefore, we have:
\begin{align}
\left|E(t_k)-E_{fw}^{App}(t_k)\right|& \le R_{max}~\delta\nonumber
\end{align}
This completes the proof.\qed

\bibliographystyle{plain}
\bibliography{ESTsurvey-1-bib}

\end{document}